%% file: main.tex
\documentclass[twocolumn]{autart} 

\usepackage{amsmath,amsfonts,amssymb,mathrsfs}
\usepackage{graphicx}
\usepackage{natbib}        
\usepackage{enumerate}
\usepackage{wrapfig}
\usepackage{subcaption}

\newtheorem{proof}{Proof}

\newtheorem{proposition}{Proposition}[section]

\newcommand{\method}[1]{{\texttt{#1}}}

\begin{document}
\begin{frontmatter}
	\title{Efficient model predictive control for nonlinear systems modelled by deep neural networks}
	
	\thanks{This work was supported by the  
		Leverhulme Trust Early Career Fellowship under Award ECF-2021-517. 
		}
		
	\author{Jianglin Lan}\ead{Jianglin.Lan@glasgow.ac.uk}
	\address{James Watt School of Engineering, University of Glasgow, Glasgow G12 8QQ, United Kingdom}
	

	\begin{abstract}
		\input{sections/abstract}
	\end{abstract}
\end{frontmatter}
	
	\section{Introduction}\label{sec:intro}
	\input{sections/introduction}
	
	\section{Problem description and preliminary}\label{sec:problemsetup}
	\input{sections/preliminary}

	\section{Model predictive control design}\label{sec:mpc tracking}
	\input{sections/tracking}

	\section{Methods for solving the MPC problem}\label{sec:method}

\input{sections/method}

	\section{Numerical example}\label{sec:simulation}
	\input{sections/simulation}

	\section{Conclusion}\label{sec:conclusion}
	\input{sections/conclusion}

	\bibliographystyle{model5-names}
	\bibliography{reference}

\end{document}

%% file: sections/abstract.tex
This paper presents a model predictive control (MPC) for dynamic systems whose nonlinearity and uncertainty are modelled by deep neural networks (NNs), under input and state constraints. 
Since the NN output contains a high-order complex nonlinearity of the system state and control input, the MPC problem is nonlinear and challenging to solve for real-time control. 
This paper proposes two types of methods for solving the MPC problem: the mixed integer programming (MIP) method which produces an exact solution to the nonlinear MPC, and linear relaxation (LR) methods which generally give suboptimal solutions but are much computationally cheaper. Extensive numerical simulation for an inverted pendulum system modelled by ReLU NNs of various sizes is used to demonstrate and compare performance of the MIP and LR methods.

%% file: sections/introduction.tex
The advancement of deep learning techniques has stimulated much interest in adopting neural networks (NNs) to power autonomous systems such as robots and self-driving cars \citep{Spielberg+19,Tang+22}.
In particular, NNs have been demonstrated to be powerful in the modelling and control of dynamic systems \cite{Moe+18}. 
This paper focuses on NN-modelled control systems, where deep NNs are 
used to model the nonlinear system dynamics and/or uncertainties. Currently, the modelling is achieved by two main types of NNs: static networks (e.g., multi-layer perception NNs) and dynamic networks (e.g., recurrent NNs, neuro-fuzzy NNs).
The highly nonlinear nature of NNs imposes big challenges on the control design of NN-modelled dynamic systems. The design becomes even more challenging when using non-differentiable NNs such as the ReLU (Rectified Linear Unit) NNs. This calls for control designs tailored to NN-modelled dynamic systems. 

There are a few existing control designs for NN-modelled dynamic systems. 
For dynamic systems modelled with a single hidden layer NN, 
an internal model control is designed in \cite{NahasSeborg92} and 
state feedback controllers are developed in \cite{Nikolakopoulou+22,DAmico+22} using the linear matrix inequality technique. An iterative linear quadratic regulator is proposed in \cite{NagariyaSaripalli20} for 
dynamic systems modelled by a deep NN.  
However, all the above designs consider only unconstrained controllers. 

To provide optimal control actions under system constraints, model predictive control (MPC) for NN-modelled dynamic systems has attracted much attention \citep{Ren+22}. This line of research is of particular interest because the combination of MPC and NN models opens the door for optimally control complex dynamic systems, which is otherwise difficult or even impossible by using first-principle models.

The MPC design for NN-modelled dynamic systems is non-trivial. In general, MPC requires solving constrained optimisation problems, but this can be burdensome for NN-modelled dynamics because the NN output normally includes high-order
nonlinearity of its input, the system state and control input. This
is particularly true for large-scale (deep) NNs with many hidden layers and neurons. 
Several methods have been adopted to solve the nonlinear MPC for NN-modelled dynamic systems, e.g., the quasi-Newton algorithm \citep{Sorensen+99},  sequential quadratic programming method \citep{Saint+91},  real-time-iteration scheme \citep{Lawrynczuk10,Patan18,Wysocki+15}, sampling method \citep{Spielberg+21,Salzmann+23}, interior point line search filter method \citep{Rokonuzzaman+21}, recurrent NN method \citep{YanWang12}, and the state observer method \citep{Bonassi+24}.
However, these methods are designed for small-scale NNs having only one or two hidden layers. 
To design MPC for dynamic systems modelled by deep NNs, sampling-based methods are developed in \cite{Williams+16,Askari+22} to obtain approximate solutions through an interative solving procedure. 

This paper presents novel efficient solving methods for MPC tracking control of deep ReLU NN-modelled nonlinear dynamic systems. 
The main contributions are as follows:
\begin{enumerate}
	\item[1)] A dual-mode MPC is proposed to ensure output of the deep NN-modelled dynamic systems track the given reference. Ways of determining the steady-state target and terminal constraint set are detailed, which are lacking in the literature \citep{Williams+16,Askari+22,Wei+22}. 
	\item[2)] Three different methods are developed to solve the nonlinear MPC problems, including the 1) MIP method by representing equivalently the NN activation function as a set of mixed integer linear constraints, 2) linear relaxation (LR) method by replacing the activation function by a triangle over-approximation of each ReLU, and 3) enhanced LR (eLR) method which refines the MPC cost function with a penalty of the deviation between the steady-state NN output value and the over-approximated NN output. 
	\item[3)] Extensive numerical simulation and ablation study are used to demonstrate and compare performance of the three solving methods, in terms of output tracking accuracy and computational efficiency. 
\end{enumerate}

The rest of the paper is structured as follows: Section \ref{sec:problemsetup} describes the problem, Section \ref{sec:mpc tracking} presents the MPC design with the solving methods in Section \ref{sec:method}, Section \ref{sec:simulation} reports the simulation results and Section \ref{sec:conclusion} draws the conclusions. 


%% file: sections/preliminary.tex
Consider a discrete-time system represented by
\begin{align}\label{eq:sys dyn}	
\begin{split}
	x(t+1) &= A x(t) + B u(t) + D f_\text{nn}(x(t),u(t)), \\
	y(t) &= C x(t),
\end{split}	
\end{align}
where $x \in \mathbb{R}^n$, $u \in \mathbb{R}^m$, and $y \in \mathbb{R}^p$ are the vectors of system state, control inputs, and measured outputs, respectively. 
$A \in \mathbb{R}^{n \times n}$, $B \in \mathbb{R}^{n \times m}$, $D \in \mathbb{R}^{n \times s}$ and $C \in \mathbb{R}^{p \times n}$ are known constant matrices. The pair $(A,B)$ is assumed to be controllable. 
$f_\text{nn}(x(t),u(t)) \in \mathbb{R}^{s}$ is a $(n+m)$-input $s$-output feedforward NN capturing the nonlinear dynamics.  

This paper considers an $(L+1)$-layer NN (with an input layer, $L-1$ hidden layers and an output layer) defined as
\begin{align}\label{eq:NN}
	\begin{split}
	z_0 &= [x(t)^\top,~ u(t)^\top]^\top, \\	
	z_i &= \phi(\hat{z}_i), ~ \hat{z}_i = W_i z_{i-1} + b_i, ~ i \in [1, L-1], \\
	f_\text{nn}(z_0) &= W_L z_{L-1} + b_L, 
	\end{split}
\end{align}
where $\hat{z}_i , z_i \in \mathbb{R}^{n_i \times 1}$ are the pre-activation and post-activation vectors at the $i$-th hidden layer with $n_i$ neurons, respectively. $W_i \in \mathbb{R}^{n_i \times n_{i-1}}$ and $b_i  \in \mathbb{R}^{n_i \times 1}$ are the $i$-th layer weight matrix and bias vector, respectively. $\phi(\cdot)$ is the ReLU function, which is among the most commonly used activation functions for NN models. The ReLU function is defined as
$\phi(\hat{z}_i) = \max\{\hat{z}_i, 0\}$, where \method{max} is applied to each element $\hat{z}_{i,j}$, $j \in [1,n_i]$, of $\hat{z}_i$. A graphic illustration of a ReLU neuron is provided in Fig. \ref{fig1}(a). 

This paper aims to design a controller $u(t)$, in the form of an MPC, such that the output $y(t)$ track a given constant reference $y_r$ whilst satisfying the constraints: 
\begin{subequations}\label{constraints}
	\begin{align}
		u \in \mathcal{U} &:= \{ u \in \mathbb{R}^m \mid \underline{u} \leq u \leq \bar{u} \}, \\
		x \in \mathcal{X} &:= \{ x \in \mathbb{R}^{n} \mid \underline{x} \leq x \leq \bar{x} \},
	\end{align}
\end{subequations}
with the known constant bounds $\underline{u}$, $\bar{u}$, $\underline{x}$ and $\bar{x}$. Since ReLU is nonlinear (piecewise and non-differentiable), the NN output $f_\text{nn}$ is a nonlinear function of its input, the system state and control input, and the order of nonlinearity increases dramatically with the hidden layers. This will impose computational challenges for solving the MPC problem.

The proposed control design will need the layer bounds of the NN, including the lower and upper bounds of the pre-activation vectors ($\hat{l}_i$ and $\hat{u}_i$, $i \in [1,L-1]$), and the output ($\underline{f}_\text{nn}$ and $\overline{f}_\text{nn}$). 
Given the constraints in \eqref{constraints}, these bounds can be computed using the interval arithmetic method as follows:
\begin{align}\label{eq:IA bounds}
		l_0 &\!=\! [\underline{x}^\top, \underline{u}^\top]^\top, ~u_0 = [\overline{x}^\top, \overline{u}^\top]^\top,\nonumber\\ 
		\hat{l}_i &\!=\! W_i^{+} l_{i-1} + W_i^{-} u_{i-1},  ~
		\hat{u}_i = W_i^{-} l_{i-1} + W_i^{+} u_{i-1}, \nonumber\\
		l_i &\!=\! \max(\hat{l}_i, 0), ~ u_i = \max(\hat{u}_i,0), ~ i \in [1,L-1], \nonumber\\
	\!\!\!\!	\underline{f}_\text{nn} &\!=\! W_L^{+} l_{L-1} \!+\! W_L^{-} u_{L-1},  
		\overline{f}_\text{nn} = W_L^{-} l_{L-1} \!+\! W_L^{+} u_{L-1},  
\end{align}
where $W_i^{+} = \max(W_i,0)$ and $W_i^{-} = \min(W_i,0)$.

%% file: sections/tracking.tex
\begin{figure*}
	\centering
	\begin{subfigure}[b]{0.3\textwidth}
		\centering
		\includegraphics[width=\textwidth]{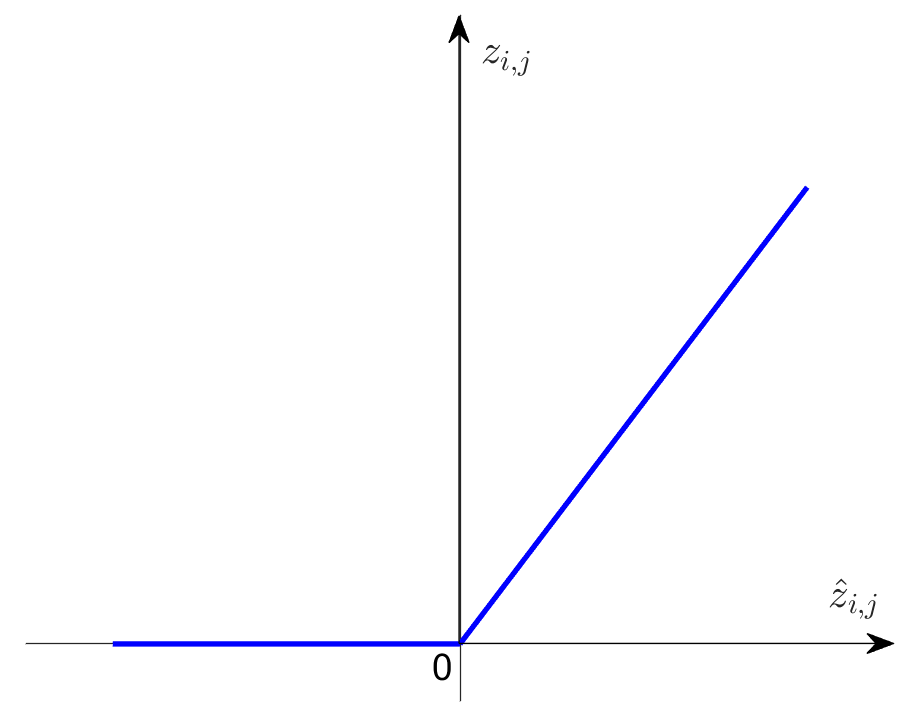}
		\caption{ReLU neuron}
		\label{fig1-1}
	\end{subfigure}
	\hfill
	\begin{subfigure}[b]{0.3\textwidth}
		\centering
		\includegraphics[width=\textwidth]{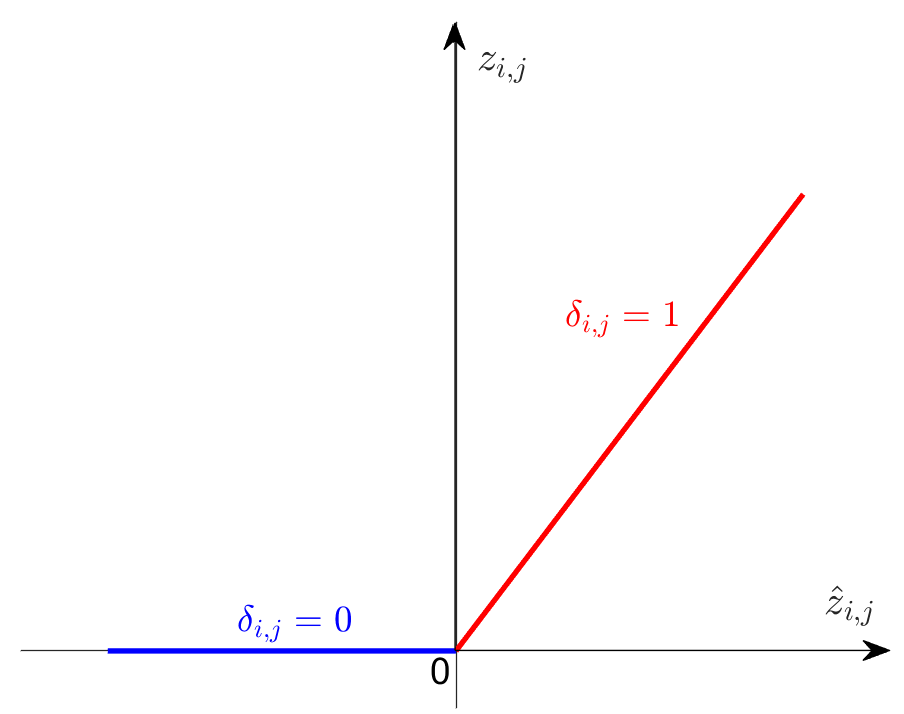}
		\caption{MIP encoding}
		\label{fig1-2}
	\end{subfigure}
	\hfill
	\begin{subfigure}[b]{0.3\textwidth}
		\centering
		\includegraphics[width=\textwidth]{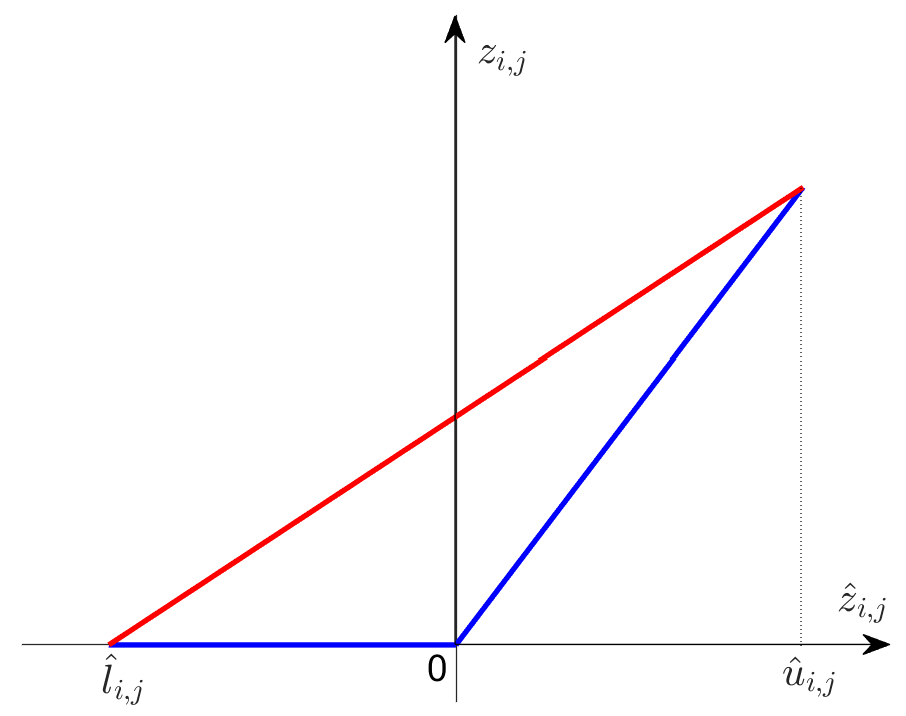}
		\caption{Linear relaxation}
		\label{fig1-3}
	\end{subfigure}
	\caption{Illustration of (a) the ReLU neuron $z_{i,j} = \phi(\hat{z}_{i,j})$ with $\hat{l}_{i,j} \leq \hat{z}_{i,j} \leq \hat{u}_{i,j}$ and its (b) MIP encoding (the neuron is inactive if $\delta_{i,j} = 0$ and active if $\delta_{i,j} = 1$) and (c) linear relaxation (the area between the blue and red lines).}
	\label{fig1}
\end{figure*}

The proposed MPC takes a dual-mode form \cite{Rossiter03} with a state feedback controller to pre-stablise the linear steady-state dyanmics of the system \eqref{eq:sys dyn}, which has not been considered in the relevant litrature. 
The overall controller is designed as
\begin{equation}\label{eq:controller form}
	u(t) = u_\text{s}(t) + u_\text{c}(t),
\end{equation}
where $u_\text{s}(t)$ and $u_\text{c}(t)$ are the steady-state feedback controller and the constraint-enforcing controller, respectively.

\textbf{Steady-state feedback controller design.}
When the system output $y(t)$ track the given output reference $y_r$ accurately, the system \eqref{eq:sys dyn} reaches the steady state $(x^*, u^*)$ that satisfies the following relation:
\begin{align}\label{eq:set point}
	\begin{split}
		x^* &= A x^* + B u^* + D f_\text{nn}(x^*,u^*), \\
		y^* &= C x^*, ~ x^* \in \mathcal{X}, ~ u^* \in \mathcal{U}.	
	\end{split}
\end{align}
In view of \eqref{eq:set point}, the target steady state $(x^*,u^*)$ can be solved from the following optimisation problem:
\begin{subequations}\label{op:set point}
	\begin{align}
		(x^*,u^*) &:=  \arg \underset{x_r,u_r}{\min} \| u_r \|^2_{R_s} \nonumber\\	
		\text{s.t.} ~& 
		\label{op:set point const1}
		\begin{bmatrix}
			I_n - A & -B	\\C & 0
		\end{bmatrix}
		\begin{bmatrix} x_r \\ u_r \end{bmatrix}
		= 
		\begin{bmatrix}
			D f_\text{nn}(x_r,u_r) \\ y_r
		\end{bmatrix}, \\
		\label{op:set point const2}
		& x_r \in \mathcal{X}, ~ u_r \in \mathcal{U},
	\end{align}
\end{subequations}
where $R_s \succ 0$ is a given weight matrix. 
This problem is nonlinear because \eqref{op:set point const1} depends on the ReLU function. It can be approximately solved by using the \method{fminsearchbnd} method \citep{Errico23}, which extends \method{fminsearch} in Matlab to incorporate constraints. This problem can be solved exactly by using the MIP method to be described in Section \ref{sec:method}, but with high computational burden. Designing more efficient methods for solving this problem exactly are left for future study.

Subtracting \eqref{eq:set point} from \eqref{eq:sys dyn} gives the tracking error system
\begin{equation}\label{eq:error sys}
	\delta x(t+1) = A \delta x(t) + B (u(t) - u^*) + D \delta f_\text{nn}(t),	
\end{equation}
where $\delta x = x - x^*$ and $\delta f_\text{nn} = f_\text{nn}(x,u) - f_\text{nn}(x^*,u^*)$.

The steady-state feedback controller $u_\text{s}(t)$ is designed as
\begin{equation}\label{eq:control law form2}
	u_\text{s}(t) = K \delta x(t) + u^*,
\end{equation} 
where $\delta x(t) = x(t) - x^*$ and $K \in \mathbb{R}^{m \times n}$ is a constant gain. In this paper $K$ is designed such that $A_s = A + B K$ is Schur stable by using the linear quadratic regulator (LQR) method \citep{AndersonMoore07} with the given state and input weights $Q \in \mathbb{R}^{n \times n} \succeq 0$ and $R \in \mathbb{R}^{m \times m} \succ 0$. 

Substituting the controller \eqref{eq:controller form} into \eqref{eq:error sys} yields
\begin{equation}\label{eq:error sys2}
	\delta x(t+1) = A_s \delta x(t) + B u_\text{c}(t) + D \delta f_\text{nn}(t),	
\end{equation}
where $u_\text{c}(t)$ is designed below through the MPC framework.

\textbf{MPC controller design.}
The MPC design needs the terminal state set $\mathcal{X}_f$ to ensure recursive feasibility. For the system \eqref{eq:sys dyn}, the set $\mathcal{X}_f$ can be constructed based on \eqref{eq:error sys2} by setting $u_\text{c}(t) = 0$, as constraints are inactive at steady state. 
Since the system matrix $A_s$ is Schur stable, the terminal set is defined as a robust positively invariant (RPI) set for \eqref{eq:error sys2}. 

Let $\mathcal{X}_\delta$ be a RPI set for the error system \eqref{eq:error sys2}  with the disturbance term $D \delta f_\text{nn}(t)$. 
Given the constraints in \eqref{constraints} and the steady-state target $(x^*,u^*)$, the intervals of $\delta x(t)$ and $\delta f_\text{nn}(t)$ can be derived as $[\underline{x} - x^*, \bar{x} - x^*]$ and
$[\underline{f}_\text{nn} - f_\text{nn}(x^*,u^*), \overline{f}_\text{nn} -f_\text{nn}(x^*,u^*)]$, where $\underline{f}_\text{nn}$ and $\overline{f}_\text{nn}$ are computed in \eqref{eq:IA bounds}.
Based on the bounds of $\delta x(t)$ and $\delta f_\text{nn}(t)$, the RPI set $\mathcal{X}_\delta$ is constructed using \citep[Algorithm 6.1]{KolmanovskyGilbert98} and given as
$\mathcal{X}_\delta = \{ \delta x(t) \in \mathbb{R}^n \mid H \delta x(t) \leq h\}$. 
Hence, the terminal set $\mathcal{X}_f$ for the system \eqref{eq:sys dyn} with the proposed controller \eqref{eq:controller form} is constructed as
\begin{equation}\label{RPI}
	\mathcal{X}_f = \{ x(t) \in \mathbb{R}^n \mid H x(t) \leq h + H x^* \}.
\end{equation}

The MPC policy at time $t$ is defined as $u_\mathrm{mpc}(t) = c^*(1|t)$ which is the first element of the optimal policy sequence $\{c(k|t)\}_{k=1}^{N}$ solved from 
\begin{subequations}\label{OP:origin mpc}
	\begin{align}
		&\hspace{2em}  \underset{\{c(k|t)\}_{k=1}^{N}, \{x(k|t)\}_{k=1}^{N+1}}{\min} V_t \nonumber\\
		\!\!\!\! &\mathrm{s.t.} ~
		\label{eq:origin mpc const1}
		 x(k+1|t) = A x(k|t) + B u(k|t) + D f_\text{nn}(k|t), \!\\
		\label{eq:origin mpc const2}
		&\hspace{1em} u(k|t) = K (x(k|t) - x^*) + u^* + c(k|t), \\
		\label{eq:origin mpc const2-2}
		&\hspace{1em} u(k|t) \in \mathcal{U}, ~x(1|t) = x(t), ~x(k|t) \in \mathcal{X},\\
		\label{eq:origin mpc const3}
		&\hspace{1em}  x(N+1|t) \in \mathcal{X}_f, \\
		\label{eq:origin mpc const4}
		&\hspace{1em} z_0(k) = [x(k|t)^\top,~ u(k|t)^\top]^\top, \\
		\label{eq:origin mpc const5}
		& \hspace{1em}z_i(k) = \phi(\hat{z}_i(k)), ~ i \in [1,L-1], \\
		\label{eq:origin mpc const6}
		&\hspace{1em} f_\text{nn}(k|t) = \hat{z}_{L-1}(k), \\
		\label{eq:origin mpc const7}
		&\hspace{1em} \underline{f}_\text{nn} \leq f_\text{nn}(k|t) \leq \overline{f}_\text{nn}, ~k \in [1,N],	 
	\end{align}
\end{subequations}
where the cost function is designed as
$V_t = \sum_{k=1}^{N} (\|x(k|t) - x^* \|^2_Q + \|u(k|t) - u^* \|^2_R ) + \|x(N+1|t) - x^* \|^2_P.$
$N$ is the prediction horizon. $Q$ and $R$ are those weight matrices used in the LQR design for $K$. $P$ is the solution to the Riccati equation $A^\top P A  - P - (A^\top P B)(R + B^\top P B)^{-1} (B^\top P A) + Q = 0$. $\mathcal{X}_f$ is the terminal constraint set to ensure recursive feasibility. $\Phi$ is a user-given weight matrix.

By implementing the proposed controller \eqref{eq:control law form2}, stability of the tracking error system under the constraints can be proved using the standard Lyapunov function methods as in \cite{Rossiter03,Rawlings+17} and the details are omitted here. 

Due to the high-order nonlinear constraints in \eqref{eq:origin mpc const5}, the optimisation problem \eqref{OP:origin mpc} is generally hard to solve. Several methods will be proposed in the next section to efficiently solve it.
Due to the use of concrete bound propagation, the interval arithmetic method in \eqref{eq:IA bounds} generates a conservative outer-approximation of $f_\text{nn}(x(t),u(t))$, resulting in a small terminal constraint set. 
Computing a tighter interval for $f_\text{nn}(x(t),u(t))$ would be useful to enlarge $\mathcal{X}_f$ and improve recursive feasibility of the MPC problem. A method to achieve this will be presented later in Section \ref{sec:method}.

%% file: sections/method.tex
This section describes methods for solving the formulated nonlinear MPC problem \eqref{OP:origin mpc}, including an exact MIP method and two linear relaxation (LR) methods.

\subsection{MIP method}
By definition, the $(i,j)$-neuron (i.e., the $j$-th ReLU neuron at the $i$-th layer) can be either active (when $z_{i,j} = \hat{z}_{i,j}$) or inactive (when $z_{i,j} = 0$).
An extra binary variable $\delta_{i,j}$ can be introduced to indicate the neuron state: When $\delta_{i,j} = 1$, the neuron is active; When $\delta_{i,j} = 0$, the neuron is inactive. Hence, each ReLU function $z_{i,j} = \phi(\hat{z}_{i,j})$ is split into two cases (as illustrated in Fig. \ref{fig1}(b)):
\begin{align}\label{eq:binary var}
	\delta_{i,j} = 1 \implies z_{i,j} = \hat{z}_{i,j}, ~
		\delta_{i,j} = 0 \implies 	z_{i,j} = 0.
\end{align}
By using \eqref{eq:binary var}, the nonlinear ReLU constraints
in~\eqref{eq:origin mpc const5} can be represented as a set of linear constraints as follows:
\begin{align}\label{MIP relaxation}
	& z_i(k) \geq 0,~ z_i(k) \geq \hat{z}_i(k),~ z_i(k) \leq \hat{z}_i(k) - \hat{l}_i \odot (1 - \delta_{k,i}),	\nonumber\\
	& z_i(k) \leq \hat{u}_i \odot \delta_{k,i},~ \delta_{k,i} \in \{0,1\}^{n_i}, ~ i \in [1,L-1], 
\end{align}
where the bounds $\hat{l}_i$ and $\hat{u}_i$ are computed in \eqref{eq:IA bounds} and $\odot$ is the element-wise product. 
It can be easily shown that \eqref{MIP relaxation} is always equivalent to \eqref{eq:binary var} providing that $\hat{l}_i \leq \hat{z}_i(k) \leq \hat{u}_i$, while conservativeness of the bounds $\hat{l}_i$ and $\hat{u}_i$ is not a matter. 

By applying \eqref{MIP relaxation}, the MPC problem \eqref{OP:origin mpc} is equivalently reformulated as the following MIP problem:
\begin{subequations}\label{OP:MIP mpc}
	\begin{align}
		&  \underset{\{c(k|t)\}_{k=1}^{N}, \{x(k|t)\}_{k=1}^{N+1},\{\delta_{k,i}\}_{k=1}^{N}}{\min} V_t \nonumber\\
		\!\!\mathrm{s.t.} ~&
		\eqref{eq:origin mpc const1} -
		\eqref{eq:origin mpc const4}, 
		\eqref{eq:origin mpc const6}, 
		\eqref{eq:origin mpc const7}, 
		\eqref{MIP relaxation}, k \in [1,N]. 
	\end{align}
\end{subequations}

In the literature, a similar MIP encoding of ReLU neurons is used to design a safe controller for NN-based dynamic systems \citep{Wei+22} and analyse the stability and feasibility of NN control systems \citep{KargLucia20}.
The MIP reformulation does not introduce any conservatism. But its computational complexity grows rapidly with the number of binary variables and the prediction horizon, which limits its scalability to large NNs and its applicability in real-time control for deep NN-modelled dynamic systems. This motivates the proposal of more computationally efficient solving methods below.

\subsection{Linear relaxation (LR) method}
The LR method consists in using a triangle relaxation to over-approximate the $(i,j)$-neuron as follows (see an illustrative example in Fig. \ref{fig1}(c)):
\begin{equation}\label{LR relaxation}
z_{i,j} \geq 0, ~z_{i,j} \geq \hat{z}_{i,j}, ~
z_{i,j} \leq a_{i,j} ( \hat{z}_{i,j}  - \hat{l}_{i,j}) + \phi(\hat{l}_{i,j}),		
\end{equation}
where $a_{i,j} = (\phi(\hat{u}_{i,j})
- \phi(\hat{l}_{i,j}))/(\hat{u}_{i,j} - \hat{l}_{i,j})$ when $\hat{l}_{i,j} \neq \hat{u}_{i,j}$ and $a_{i,j} = 0$ otherwise. Note that an over-approximation is necessary to ensure that if the computed MPC policy satisfies the state and input constraints, then it is still the case when applying the policy to the original system model. This is, however, not achievable by using the sampling-based methods \citep{Williams+16,Askari+22} or using approximation of the ReLU such as the polynomial approximation \citep{Kochdumper+22}.

Replacing the ReLU constraints \eqref{eq:origin mpc const5} with \eqref{LR relaxation} yields the LR method:
\begin{subequations}\label{OP:LR mpc}
	\begin{align}
		&  \underset{\{c(k|t)\}_{k=1}^{N}, \{x(k|t)\}_{k=1}^{N+1}}{\min} V_t \nonumber\\
		\!\!\mathrm{s.t.} ~&
		\label{eq:LR mpc const1-4}
		\eqref{eq:origin mpc const1} - 
		\eqref{eq:origin mpc const4},
		\eqref{eq:origin mpc const6}, 
		\eqref{eq:origin mpc const7},\\
		\label{eq:LR mpc const5}
		& \eqref{LR relaxation}, j \in [1,n_i], i \in [1,L-1], k \in [1,N-1]. 
	\end{align}
\end{subequations}
The problem \eqref{OP:LR mpc} is relatively easier to solve than the MIP problem \eqref{OP:MIP mpc}, but there is usually a large relaxation gap, especially for large-scale NNs, which leads to undesirable output tracking performance. A method is further developed below to improve the solution quality of the LR method.

\subsection{Enhanced linear relaxation (eLR) method}
Due to the use of over-approximation of ReLU in the LR method, the computed MPC cannot ensure the system reach the target steady state, leading to inaccurate output tracking. To improve the tracking accuracy, 
the cost function is redesigned as
\begin{align}\label{eq:new cost function}
	& J_t = \sum_{k=1}^{N} (\|x(k|t) - x^* \|^2_Q + \|u(k|t) - u^* \|^2_R \nonumber\\
	&+ \| f_\text{nn}(k|t) - f_\text{nn}(x^*, u^*) \|^2_\Phi ) + \|x(N+1|t) - x^* \|^2_P,
\end{align}
where the term $\| f_\text{nn}(k|t) - f_\text{nn}(x^*, u^*) \|^2_\Phi$ with a user-given weight matrix $\Phi$ is introduced to minimise the deviation of the over-approximated NN output $f_\text{nn}(k|t)$ from its steady-state value $f_\text{nn}(x^*, u^*)$. By minimising this new cost function, the obtained MPC policy can minimise the total deviation of the system from its steady-state dynamics in \eqref{eq:set point}.
Due to introduction of the new term, it holds that $J_t \geq V_t$.

Applying the new cost function in \eqref{eq:new cost function} to \eqref{OP:LR mpc} results in the enhanced LR (eLR) method:
\begin{subequations}\label{OP:enhanced LR mpc}
\begin{align}
	& \underset{\{c(k|t)\}_{k=1}^{N}, \{x(k|t)\}_{k=1}^{N+1}}{\min} J_t \nonumber\\
	\!\!\mathrm{s.t.} ~&
	\label{eq:eLR mpc const1-4}
	\eqref{eq:origin mpc const1} - \eqref{eq:origin mpc const4},~	\eqref{eq:origin mpc const6}, ~\eqref{eq:origin mpc const7},\\
	\label{eq:eLR mpc const5}
	& \eqref{LR relaxation}, j \in [1,n_i], i \in [1,L-1], k \in [1,N-1].	
\end{align}		
\end{subequations}
Note that when the optimal solution to \eqref{OP:enhanced LR mpc} satisfies $J_t^* = 0$, the obtained MPC ensures that the system reaches the  target steady-state and accurate output tracking.

\subsection{Discussions}
Proposition \ref{proposition:comparison} provides a comparison of the solution quality (optimality or tracking accuracy) $S_\text{method}$ of the MIP, LR, and eLR methods against the original MPC \eqref{OP:origin mpc}.
\begin{proposition}\label{proposition:comparison}
	The solution quality of the proposed methods satisfy: 
	$S_\text{LR} \leq S_\text{eLR} \leq S_\text{MIP} = S_\text{origin}.$
\end{proposition}

The LR and enhanced LR methods are of the same size and will have roughly the same computational complexity, while the MIP method is generally much more computationally expensive. A further demonstration of this is to be through simulation in Section \ref{sec:simulation}.
 
Computational complexity of the proposed solving methods can be reduced via directly encoding the stable (either inactive or active) neurons using equality constraints. 
For each $(i,j)$-neuron, the pre-activation bounds $\hat{l}_{i,j}$ and $\hat{u}_{i,j}$ can be used to identity its status as follows: 1) If $\hat{u}_{i,j} \leq 0$, the neuron is known to be strictly inactive and thus $z_{i,j} = 0$, and 2) If $\hat{l}_{i,j} \geq 0$, the neuron is known to be strictly active and thus $z_{i,j} = \hat{z}_{i,j}$. 
Applying these to the proposed methods can reduce the required numbers of binary variables or triangle relaxation. This can then improve the computational efficiency without compromising the solution quality. It is also worth noting that more stable neurons can be identified if less conservative pre-activation bounds are used. The symbolic bound propagation method \citep{HenriksenLomuscio20} can be an efficient way to achieve this and it will be explored in the future work. 

This paper focuses on developing efficient algorithm for solving the MPC problem, so the NN $f_\text{nn}$ is assumed to capture the system nonlinear dynamics and uncertainty. If a separate term is used to lump the total effects of uncertainty and diturbance, the proposed MPC can be extended as robust MPC based on the concept of tube-MPC \citep{Langson+04}, in which the proposed solving algorithms will still be applicable.

%% file: sections/simulation.tex
Consider an inverted pendulum system
\begin{align}\label{sim:sys1}
	x_1(t+1) &= x_1(t) + t_s x_2(t), \nonumber\\
	x_2(t+1) &= x_2(t) + \frac{t_s g\sin(x_1(t))}{l} - \frac{t_s c x_2(t)}{m l^2}  + \frac{t_s u(t)}{m l^2}, \nonumber\\
	y(t) &= x_1(t), \nonumber
\end{align}
where $x_1(t)$ is the angular displacement, $x_2(t)$ is the angular velocity, $t_s$ is the sampling time, $m$ is the mass of pendulum, $l$ is the distance from pivot to centre of mass of the pendulum, $c$ is the rotational friction coefficient, and $g$ is the gravitational constant. 
The parameters used in the simulation are $t_s = 0.1~\mathrm{s}$, $m = 1$, $l = 1$, $g = 9.8$ and $c = 0.01$. The input and state constraints are: $|u(t)| \leq 3, ~|x(t)| \leq [\pi/2,~5]^\top$.

Suppose the value of the rotational friction coefficient $c$ is unknown for the control design. The pendulum system is reformulated as
\begin{equation}\label{sim:sys2}
	\begin{split}
		x(t+1) &= A x(t) + B u(t) + D f(x(t)), \\
		y(t) &= C x(t),
	\end{split}	
\end{equation}
where
$
A = 
\begin{bmatrix}
1 & t_s \\ 0 & 1	
\end{bmatrix}$, 
$B = \begin{bmatrix}
	0 \\ \frac{t_s}{m l^2} 
\end{bmatrix}$, 
$D = \begin{bmatrix}
	0 \\ \frac{t_s}{l} 
\end{bmatrix}$, 
$C = [1 ~ 0]$ and
$f(x(t)) = g\sin(x_1(t)) - (c x_2(t))/(m l). 
$

A three-input one-output NN $f_\text{nn}(x(t),u(t))$ is trained to model the unknown nonlinear dynamics $f(x(t))$. Then the pendulum system can be represented in the form of \eqref{eq:sys dyn}. 
The simulations are conducted in Matlab on a Linux machine with an AMD Ryzen Threadripper PRO 5955WX 16-cores CPU and 128 GB RAM. The optimisation problems are modelled using YALMIP \citep{lofberg2004yalmip} and solved by Gurobi \citep{gurobi}. The weight matrices used in the control design are: $R_s = 1$, $Q = [1.0\mathrm{e}5, 0; 0, 1.0\mathrm{e}2]$, $R = 1$, and $\Phi = 100$. The initial state is $x(0) = [0;0]$ and the output reference is $y_r = \pi/5$.

\begin{figure}[t]
	\centering
	\includegraphics[width=\columnwidth]{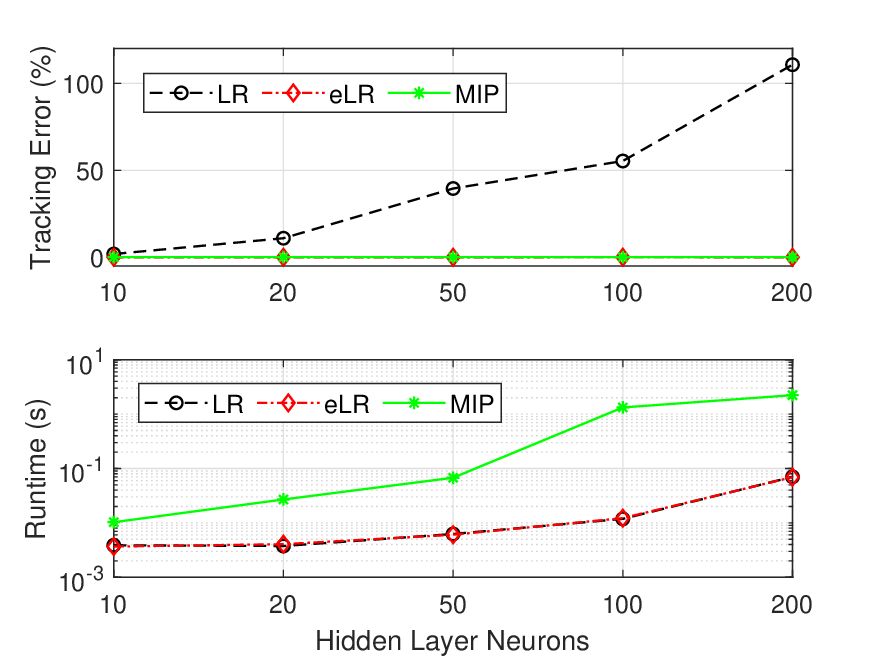}
	\vspace{-1.5em} 
	\caption{Tracking performance and runtime for 3-layer NNs with different number of neurons at the hidden layer.}
	\label{fig2}
\end{figure}

\begin{figure}[t]
	\centering
	\includegraphics[width=\columnwidth]{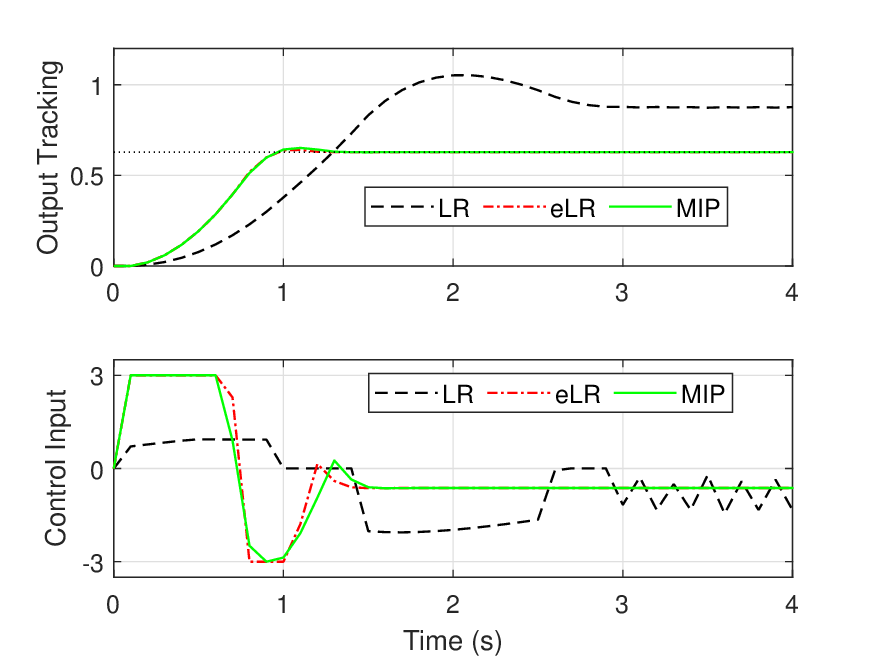}
	\vspace{-1.5em} 
	\caption{Example output tracking and control input for the 3-layer NN with 50 neurons in Case 1.}
	\label{fig3}
\end{figure}

\begin{figure}[t]
	\centering
	\includegraphics[width=\columnwidth]{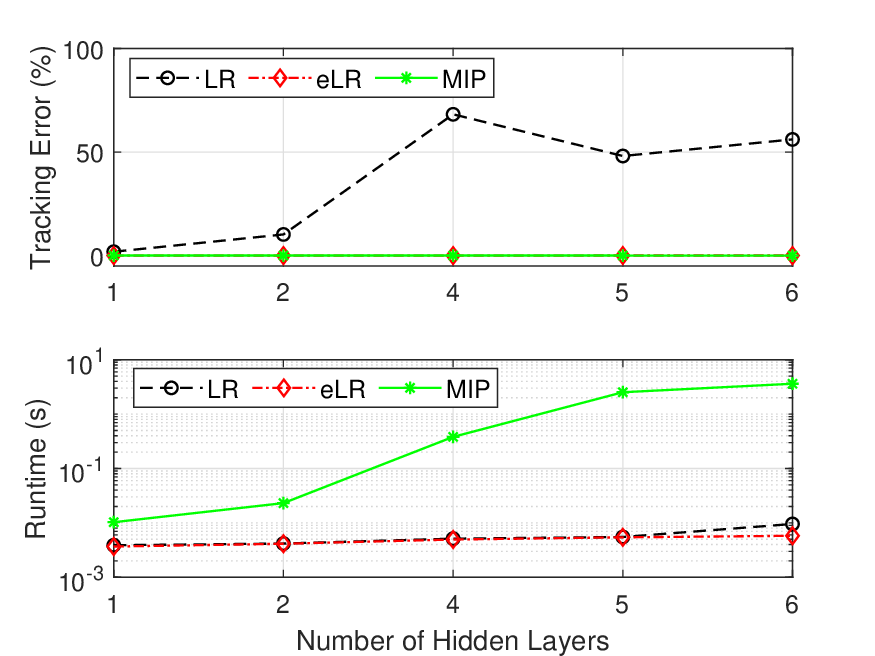}
	\vspace{-1.5em} 
	\caption{Tracking performance and runtime for NNs with different number of hidden layers each having 10 neurons.}
	\label{fig4}
\end{figure}

\begin{figure}[t]
	\centering
	\includegraphics[width=\columnwidth]{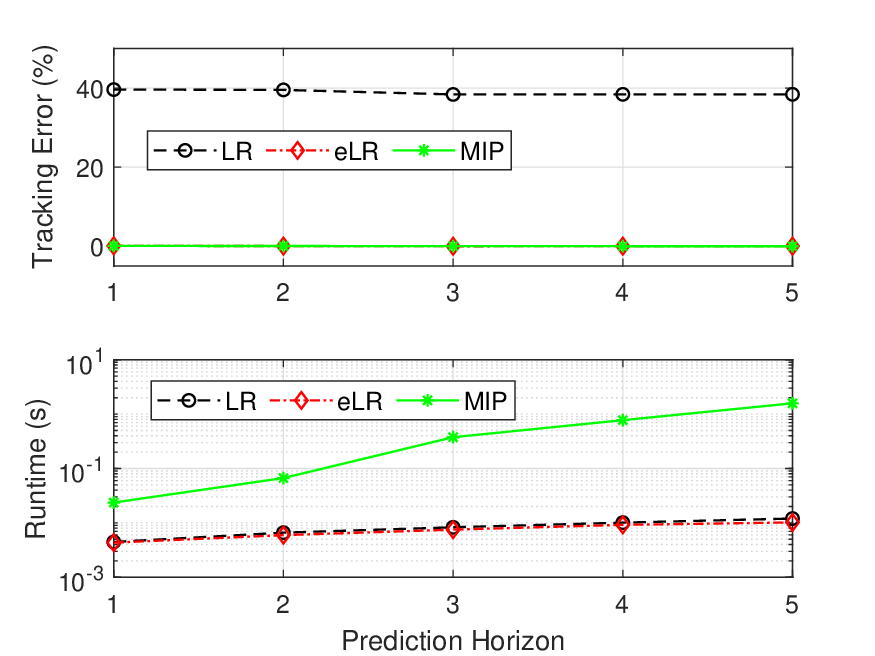}
	\vspace{-1.5em} 
	\caption{Tracking performance and runtime for 3-layer NNs with 50 neurons under different prediction horizons.}
	\label{fig5}
\end{figure}

Comparison is made for the MIP method in \eqref{OP:MIP mpc}, the LR method in \eqref{OP:LR mpc}, and the eLR method in \eqref{OP:enhanced LR mpc}. 
The performance metrics to compare are the steady-state relative tracking error $|y - y_r|/y_r \times 100$ and the maximum runtime for solving the MPC problems at each sampling step. 
Three simulation cases are used to make a comprehensive comparison:
\begin{itemize}
	\item \textbf{Case 1: Network width test.} This case simulates 3-layer NNs where the number of neurons at the only hidden layer is 10, 20, 50, 100, or 200, repectively. The prediction horizon is $N=1$. The results are depicted in Fig. \ref{fig2} and an example of the outputs and control inputs for the 50 neurons NN is provided in Fig. \ref{fig3}.
	\item \textbf{Case 2: Network depth test.} This case uses NNs having 1, 2, 4 or 5 hidden layers with 10 neurons at each hidden layer. The prediction horizon is $N=1$. The results are reported in Fig. \ref{fig4}. 
	\item \textbf{Case 3: Prediction horizon test.} 
	This case considers a 3-layer NN with 50 neurons. The prediction horizon is $N=2, 3, 4, 5$, respectively. The results are reported in Fig. \ref{fig5}. 
\end{itemize}

In all the three cases, the tracking accuracy from the best to the worst are: MIP, eLR, and LR.
Moreover, the relative tracking errors of the LR method grow rapidly with increase in the network width (i.e., the number of neurons) and depth (i.e., the number of hidden layers). But the tracking errors of the MIP and eLR methods are almost zero in all simulations. 

As expected, the relaxation methods, LR and eLR methods, need around the same runtime in all the simulations, while the MIP method is much more computationally expensive. 
Moreover, the runtime of MIP grows much faster than the other methods when the NNs become larger (with more neurons or hidden layers) or the prediction horizon becomes longer. The runtime of MIP method exceeds the sampling time 0.1 s when the 3-layer NNs have more than 50 neurons per layer (as shown in Fig. \ref{fig2}), when the NNs have more than 2 hidden layers (as shown in Fig. \ref{fig4}), or when the prediction horizon is longer than 2 (as shown in Fig. \ref{fig5}). However, the runtime for the relaxation methods remains lower than the sampling time across all the NNs and simulations.

%% file: sections/conclusion.tex
A dual-mode MPC is designed to ensure output set point tracking of NN-modelled dynamic systems in the presence of state and input constraints. 
Both exact MIP method and linear relaxation (LR) methods are proposed to solve the formulated nonlinear MPC problem. Theoretic analysis and simulation evaluation show that the enhanced LR can enjoy both the benefits of low computational burden from the LR method and accurate output tracking from the MIP method. This suggests the enhanced LR method is more suitable for real-time control of deep NN-modelled dynamic systems.  
Future work will be extending the proposed control design for NNs with other types of activation functions.